\RequirePackage{snapshot}

\documentclass[utf8]{FrontiersinHarvard} 

\usepackage{url,hyperref,lineno,microtype,subcaption}
\usepackage[onehalfspacing]{setspace}
\usepackage{comment}
\usepackage{multirow}
\usepackage{gensymb}
\usepackage{wrapfig}
\usepackage{float}

\newcommand{\argmin}[1]{\underset{#1}{\operatorname{argmin}}\;}

\def\keyFont{\fontsize{8}{11}\helveticabold }
\def\firstAuthorLast{Shemonti {et~al.}} 
\def\Authors{Abida Sanjana Shemonti\,$^{1}$, Emanuele Plebani\,$^{2}$, Natalia P. Biscola\,$^{3}$, Deborah M. Jaffey\,$^{7}$, Leif A. Havton\,$^{3,4,5}$, Janet R. Keast\,$^{6}$, Alex Pothen\,$^{1}$, M. Murat Dundar\,$^{2}$, Terry Powley\,$^{7}$ and Bartek Rajwa\,$^{8*}$}

\def\Address{$^{1}$Purdue University, Dept. of Computer Science, West Lafayette, IN, USA \\
$^{2}$Indiana University - Purdue University Indianapolis, Dept. of Computer \& Information Sciences, Indianapolis, IN, USA\\
$^{3}$Icahn School of Medicine at Mount Sinai, Dept. of Neurology, New York, NY, USA\\
$^{4}$Icahn School of Medicine at Mount Sinai, Dept. of Neuroscience, New York, NY, USA\\
$^{5}$James J. Peters Department of Veterans Affairs Medical Center, Bronx, NY, USA\\
$^{6}$University of Melbourne, Dept. of Anatomy and Physiology, Melbourne, Vic, Australia\\
$^{7}$Purdue University, Dept. of Psychological Sciences, West Lafayette, IN, USA\\
$^{8}$Bindley Bioscience Center, Purdue University, West Lafayette, IN, USA
}

\def\corrAuthor{Bartek Rajwa}\def\corrEmail{brajwa@purdue.edu}

\begin{document}
\onecolumn
\firstpage{1}

\title[Methodology for quantifying the spatial arrangements of axons]{A novel statistical methodology for quantifying the spatial arrangements of axons in peripheral nerves} 

\author[\firstAuthorLast ]{\Authors} 
\address{} 
\correspondance{} 

\extraAuth{}

\maketitle

\begin{abstract}

A thorough understanding of the neuroanatomy of peripheral nerves is required for a better insight into their function and the development of neuromodulation tools and strategies. In biophysical modeling, it is commonly assumed that the complex spatial arrangement of myelinated and unmyelinated axons in peripheral nerves is random, however, in reality the axonal organization is inhomogeneous and anisotropic. Present quantitative neuroanatomy methods analyze peripheral nerves in terms of the number of axons and the morphometric characteristics of the axons, such as area and diameter. In this study, we employed spatial statistics and point process models to describe the spatial arrangement of axons and Sinkhorn distances to compute the similarities between these arrangements (in terms of first- and second-order statistics) in various vagus and pelvic nerve cross-sections. We utilized high-resolution TEM images that have been segmented using a custom-built high-throughput deep learning system based on a highly modified U-Net architecture. Our findings show a novel and innovative approach to quantifying similarities between spatial point patterns using metrics derived from the solution to the optimal transport problem. We also present a generalizable pipeline for quantitative analysis of peripheral nerve architecture. Our data demonstrate differences between male- and female-originating samples and similarities between the pelvic and abdominal vagus nerves.     
\tiny
 \keyFont{ \section{Keywords: Peripheral nervous system, Neuroanatomy, Neuromodulation, Spatial point process, Optimal transport problem, Sinkhorn distance} } 
 
\end{abstract}

\section{Introduction}

Understanding the functionalities of the peripheral nerves and developing neuromodulation tools require an in-depth quantitative characterization of the anatomy of the nerves. A large portion of the quantitative neuroanatomical studies focus on counting and comparing the number of myelinated and unmyelinated axons in the peripheral nerves in different animals \citep{hoffman_numbers_1961, krous_developmental_1985, asala_electron_1986, prechtl_fiber_1990, pereyra_development_1992, soltanpour_preservation_1996, safi_myelinated_2016}. There are studies on analyzing the changes in the number of myelinated and unmyelinated axons as the function of animals' age \citep{krous_developmental_1985, pereyra_development_1992, soltanpour_preservation_1996}. The morphometric characteristics of the axons, such as area of axon cross-section, diameter, myelin thickness, are also well-developed and helpful for estimating electrode distances for neuromodulation purposes \citep{asala_electron_1986, prechtl_fiber_1990, walter_differential_2019, pelot_quantified_2020, havton_human_2021, settell_vivo_2021}.

The vagus is a complex, multi-functional peripheral nerve of the autonomic nervous system, containing both sensory and motor axons that regulate a wide variety of functions \citep{camara_chapter_2015, breit2018vagus}. These include regulation of the heart, respiratory tract, and many areas of the gastrointestinal system, influencing motility, secretions, and communication with the immune system. This breadth of activity and its bidirectional connectivity with the central nervous system have led to the vagus becoming a promising target for bioelectric medicine, through development of specific protocols for vagal nerve stimulation (VNS) \citep{bonaz_vagus_2017, bonaz2017vagus, Horn2019}.

In addition to providing an alternative therapeutic approach for drug-resistant clinical conditions within organs, the vagal afferent connections to the brain provide opportunities for novel therapies directed to various psychiatric disorders. To improve the efficacy and specificity of VNS for each type of clinical condition, a greater understanding of the intra-vagal neural elements relating to each organ system is required \citep{howland_vagus_2014, thompson_avoiding_2019, thompson_organotopic_2022}, as demonstrated by a recent study showing that fascicle-selective stimulation can reduce off-target effects of VNS \citep{thompson_organotopic_2022}. This includes understanding the spatial organization of different functional classes of axons within and between fascicles. This spatial organization has not been investigated in depth within the visceral nervous system.  
		
We have begun to address this knowledge gap using an extensive dataset of transmission electron microscopy (TEM) images derived from multiple cross-sections of the rat vagus. We have included in our study additional TEM images from the rat pelvic nerve, another multi-functional major nerve of the autonomic nervous system that supplies sensory and motor axons to the urogenital organs and lower bowel. Both TEM data sets have been published through the SPARC Portal (RRID: SCR\_017041) under a CC-BY 4.0 license \citep{plebani_high-throughput_2022}.  	
		


Cross-sections of large peripheral nerves reveal a variety of components (myelinated and unmyelinated axons, Schwann cells), high-order structures (fascicles and Remak bundles), and raise questions regarding the spatial arrangement of these components, similarities between multiple arrangements, and their relationship to various biological factors including age, sex, and diseases. This study focuses on the unmyelinated axons segmented utilizing our high-throughput deep learning model \citep{plebani_high-throughput_2022}. We aim to define a notion of similarity (or dissimilarity) between the spatial arrangements of the unmyelinated axons and quantify the distances between them. We resort to spatial point patterns to represent the image data conveniently for analyzing the axons' spatial organization. We use the centroids of the segmented axons to construct spatial point patterns. We consider spatial inhomogeneity and anisotropy to be the spatial features to represent the spatial arrangement of the axons, and be used for quantification. A known way to get an intuitive sense of spatial inhomogeneity (inhibition and/or attraction between points) and anisotropy of a point pattern is to investigate its second-order statistics \citep{Ripley1976, Ripley1977, Dixon2014}. 

Since Ripley’s summary of spatial statistical methods in 1977 the techniques for spatial pattern analysis have been occasionally employed in neuroscience, often by statisticians who saw the extraordinary complexity of neuroanatomical patterns to be a perfect demonstration of the spatial statistics inference ability \citep{Bjaalie1991, Diggle1991,Prodanov2007,Jafari-Mamaghani2010,Waller2011}. 

However, the standard spatial statistical measures are not well suited for quantifying a definite distance between different non-random patterns. Therefore, we propose a method that involves computing the local second-order spatial statistics for the nerve fascicles to capture their spatial arrangement and utilizing a revised optimal transport distance (Sinkhorn distance) to measure similarities between the second-order spatial statistics of every pair of nerve cross-sections. We visualize the resulting Sinkhorn distance matrix in a new metric space using multi-dimensional scaling that helps interpret the similarity (dissimilarity) of the spatial features in the nerve cross-sections. Our concept of Sinkorn distance embedding was influenced by related work on optimal transport-based morphometry applications in cell biology \citep{Wang2013c, Basu2014}.

In addition to addressing a neuroscientific problem of quantifying the vagus nerve anatomy with computer science tools, we intend to bring together a variety of approaches from various computer science domains to extend the toolkit for point-pattern comparisons in biology. Utilizing the optimal transport framework to establish a quantitative measure for spatial point patterns and advance the workflow of quantitative analysis of peripheral nerve architecture, our method could be used beyond neuroscience. We believe that our findings contribute to the establishment of spatially selective stimulation of nerve axons to improve the efficacy of VNS. 

\begin{figure}[t]
    \centering
    \includegraphics{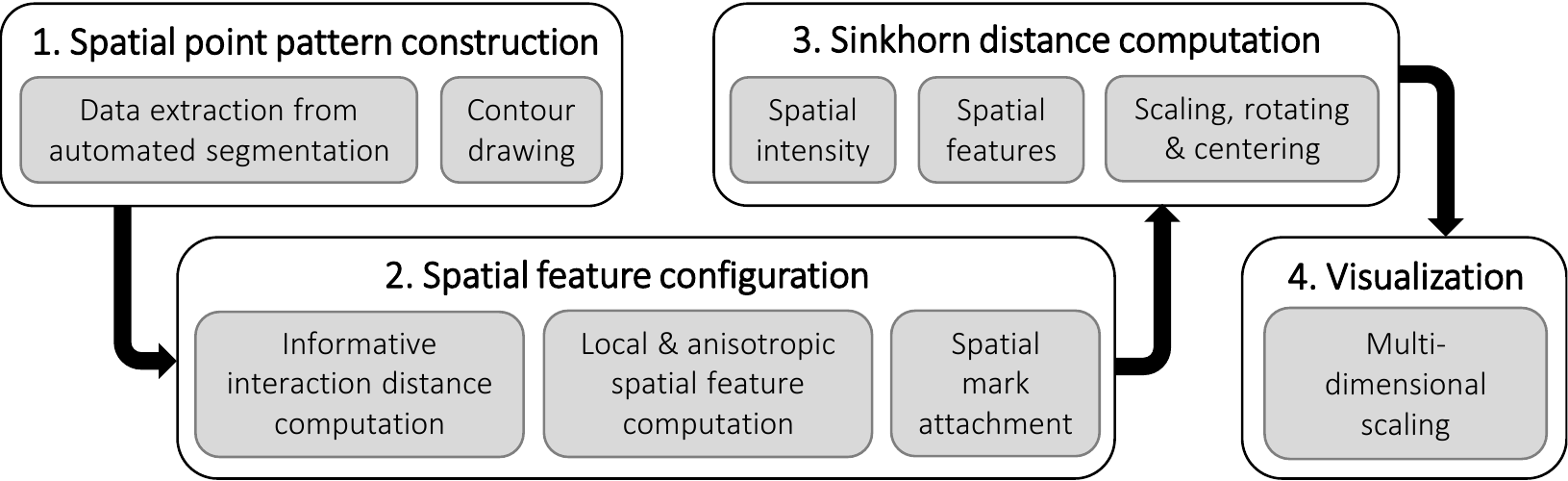}
    \caption{The pipeline for the quantitative analysis of the spatial arrangement of axons in the peripheral nerve cross-sections.}
    \label{fig:pipeline}
\end{figure}

Figure~\ref{fig:pipeline} depicts a high-level overview of the steps in the quantitative analysis of the spatial arrangement of axons in the peripheral nerve architecture presented in this paper. We explain each component of the pipeline in the following sections. The biological data and the data preprocessing steps are described in Section~\ref{sec:data}. The basics of spatial point pattern, spatial statistics, and optimal transport framework are introduced in Section~\ref{sec:spp} and Section ~\ref{sec:ot}, respectively. We provide details on the experimental setup and results in Section~\ref{sec:experiments}, which cover steps $2$, $3$, and $4$ of the computational pipeline. We discuss the results of the empirical study in Section~\ref{sec:discussion} before concluding. 
\section{Materials and Methods}

\subsection{Biological data and automated segmentation}\label{sec:data}

\begin{figure}[t]
    \centering
    \includegraphics[width=\textwidth]{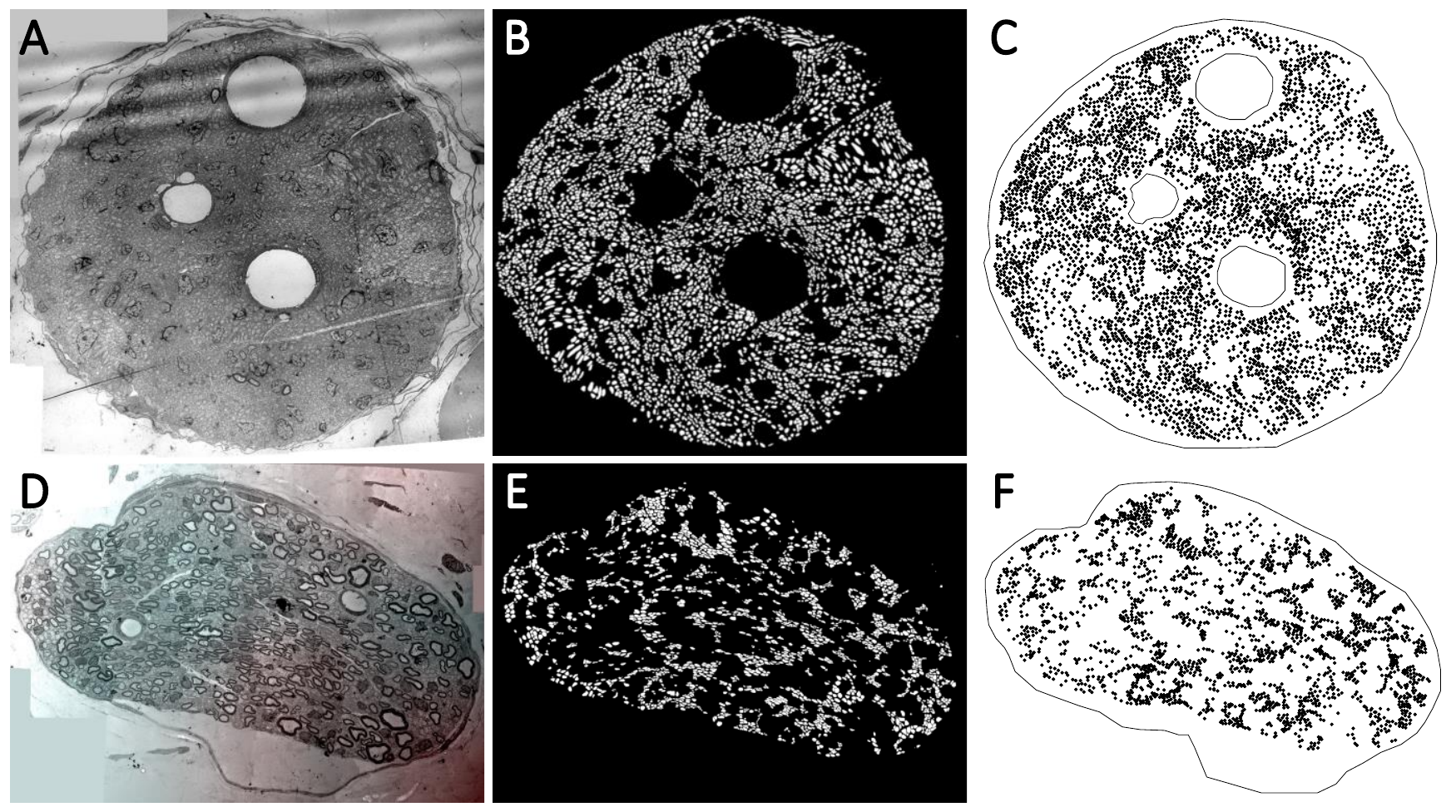}
    \caption{(A, D) The transmission electron microscopy (TEM) images of the nerve cross-section of Image 15 (vagus) and Image 29 (pelvic) listed in Table~\ref{tab:dataset} respectively. The visible void spaces in the nerve cross-sections are blood vessels. The tiny light grey regions without any border are the unmyelinated axons. The myelinated axons have slightly darker grey borders. (B, E) The automated segmentation of the unmyelinated axons (the white regions) in the nerve cross-sections. (C, F) The spatial point patterns constructed with the centroid locations (the black circles) of the segmented unmyelinated axons.}
    \label{fig:fascicle_segment_pp}
\end{figure}

Although the vagus nerve anatomy was the primary motivation behind this study, we use the TEM images of the vagus and pelvic nerve cross-sections in rats for comparisons. A list of the TEM images used in this study is shown in Table~\ref{tab:dataset}. The protocols and techniques followed for nerve sample collection, processing, and imaging are documented in our previous work \citep{plebani_high-throughput_2022}. The data is publicly available via NIH-supported SPARC Pennsieve database \citep{high_throughput_2022_dataset_pennsieve}.
Briefly, the unmyelinated axons in some of these TEM images were manually annotated and used as labeled data to train, validate, test, and evaluate an automated segmentation model based on the U-Net architecture \citep{ronneberger_u-net_2015, plebani_high-throughput_2022}. The segmentation model is a U-Net with four stages: the convolutional layers have a batch normalization layer followed by a ReLU activation layer, and the bottleneck stage has extra dropout layers between convolutions \citep{plebani_high-throughput_2022}. The model classifies the TEM image pixels as one of the three following classes: (a) \textit{fiber} if it is inside an unmyelinated axon, (b) \textit{border} if it is in a boundary region between an axon and the rest of the image defined by the outer edge of each axon, and (c) \textit{background}. An updated version of the model\footnote{at \url{https://github.com/Banus/umf_unet}} was used here to segment the unmyelinated axons. The resulting axon counts are listed in Table~\ref{tab:dataset}. We used the open-source image processing package Fiji \citep{schindelin_fiji_2012} to extract the centroid coordinates of the segmented unmyelinated axons and the functions in R packages to establish the outer boundaries and inner void spaces of the nerve cross-sections, to construct the spatial point patterns. Images 15 (vagus) and 29 (pelvic) listed in Table~\ref{tab:dataset} are shown in Figure~\ref{fig:fascicle_segment_pp}A and Figure~\ref{fig:fascicle_segment_pp}D respectively, along with their corresponding automated segmentations and spatial point patterns.

\setlength{\tabcolsep}{10pt}
\renewcommand{\arraystretch}{1.2}
\begin{table}[t]
    \centering
    \caption{The list of the TEM images of vagus and pelvic nerve cross-sections in rats used in this study. The vagus and the pelvic nerve cross-sections are collected from the following nerve locations - CT: Cervical Trunk, AVAT: Abdominal Vagus Anterior Trunk, AVPT: Abdominal Vagus Posterior Trunk, AVAG: Abdominal Vagus Anterior Gastric (Ventral Gastric Branch), PG: Pelvic Ganglion. The information regarding the Images 12-29 were published in \cite{plebani_high-throughput_2022}.}
    \begin{tabular}{c c c c c c c}
    \\ \hline
        \multirow{2}{*}{\textbf{Image ID}} & \textbf{Image size}  & \textbf{Resolution} & \multirow{2}{*}{\textbf{Nerve}} & \multirow{2}{*}{\textbf{Location}} & \multirow{2}{*}{\textbf{Sex}} & \textbf{\# Segmented} \\
        & \textbf{(pixel$\times$pixel)} &  \textbf{(nm/pixel)} & & & & \textbf{axons}\\
    \hline
         1 & 2994$\times$2497 & 11.9 & Vagus & Right CT & F & 183\\
         2 & 10624$\times$6686 & 11.9 & Vagus & Right CT & F & 5020\\
         3 & 21005$\times$22847 & 11.9 & Vagus & Right CT & F & 13375\\
         4 & 7707$\times$7978 & 11.9 & Vagus & AVAT & F & 4538\\
         5 & 9633$\times$15046 & 11.9 & Vagus & AVPT & F & 10328\\
         6 & 13120$\times$14400 & 11.9 & Vagus & AVAG & F & 6566\\
         7 & 19921$\times$9680 & 8.7 & Vagus & AVPT & F & 8980\\
         8 & 5175$\times$3784 & 11.9& Vagus & AVAG & F & 407\\
         9 &  12328$\times$9692 & 13.7 & Vagus & Right CT & M & 7647\\
         10 & 6794$\times$5472 & 13.7 & Vagus & Right CT & M & 871\\
         11 & 5262$\times$7111 & 13.7 & Vagus & AVPT & M & 9109\\
         12 & 24746$\times$20682 & 11.9 & Vagus & Right CT & F & 12992\\
         13 & 20372$\times$27269 & 11.9 & Vagus & Left CT & F & 14155\\
         14 & 7953$\times$5781 & 11.9 & Vagus & AVAG & F & 1698\\
         15 & 8446$\times$7258 & 13.7 & Vagus & AVAG & M & 4938\\
         16  & 4128$\times$4068 & 13.7 & Vagus & AVAG & M & 1061\\
         17 & 9935$\times$8870 & 13.7 & Vagus & AVAG & M & 4654\\
         18 & 5521$\times$4971 & 13.7 & Vagus & AVAG & M & 1409\\
         19 & 8633$\times$8866 & 11.9 & Pelvic & $\leq$2mm from PG & M & 1663\\
         20 & 3891$\times$3334 & 11.9 & Pelvic & $\leq$2mm from PG & M & 297\\
         21 & 2754$\times$2958 & 11.9 & Pelvic & $\leq$2mm from PG & M & 209\\
         22 & 3357$\times$3823 & 11.9 & Pelvic & $\leq$2mm from PG & M & 303\\
         23 & 4419$\times$5701 & 11.9 & Pelvic & $\leq$2mm from PG & M & 608\\
         24 & 2804$\times$4221 & 11.9 & Pelvic & $\leq$2mm from PG & M & 350\\
         25 & 5064$\times$7207 & 11.9 & Pelvic & $\leq$2mm from PG & M & 652\\
         26 & 5869$\times$6268 & 11.9 & Pelvic & $\leq$2mm from PG & M & 990\\
         27 & 7941$\times$6372 & 11.9 & Pelvic & $\leq$2mm from PG & M & 1372\\
         28 & 4028$\times$3513 & 11.9 & Pelvic & $\leq$2mm from PG & M & 460\\
         29 & 11129$\times$7962 & 11.9 & Pelvic & $\leq$2mm from PG & M & 2363\\
    \hline
    \end{tabular}
    \label{tab:dataset}
\end{table}

Before reporting the experimental details, we provide a brief overview of spatial point patterns, spatial statistics concepts, and optimal transport framework in the following two subsections.

\subsection{Spatial point patterns and spatial statistics}\label{sec:spp}

A \emph{spatial point pattern} (SPP) is a set of spatial locations associated with entities of interest in 2-D or 3-D space, encompassed by an observation window \citep{ moller2003statistical, Stoyan2006, Jafari-Mamaghani2010, baddeley2015spatial}. The analysis of SPP aims to study the spatial arrangement of the points in an SPP and establish trends to describe the point pattern. Two fundamental descriptive characteristics of an SPP are \emph{intensity} and \emph{interaction}. The intensity ($\rho$ or $\rho(u)$) of a point pattern, a first-order statistics, is the average number of points per unit area, and it could be uniform across the observation window (homogeneous), or it could vary according to an intensity function (inhomogeneous). A point pattern's intensity is usually denoted by $\lambda$ in literature, but we use $\rho$ to avoid confusion with another notation related to the optimal transport problem. The interaction is associated with a distance ($r$) and describes the influence the points have on their neighbors within $r$ radius. The interaction is termed complete spatial randomness (CSR) if the points are independent. The points can exhibit positive interaction (spatial attraction), negative interaction (spatial inhibition), or a combination of both. 

It is common practice to use second-order statistics such as Besag's centered $L$-function, which is a transformation of Ripley's $K$-function \citep{Ripley1976, Ripley1977, Besag1977}, to investigate the interaction in point patterns. Let $\mathbf{X}$ be a point pattern and $t(u,r,\mathbf{X})$ be the number of points in $\mathbf{X}$ which lie within distance $r$ of the location $u$. Assuming $\mathbf{X}$ is a homogeneous point pattern with intensity $\rho$, the number of points within distance $r$ of a specific point is represented by $\rho K(r)$ \citep{Ripley1976}.  
\begin{equation}\label{eq:k-l-funct}
    \begin{gathered}
        \text{Ripley’s $K$-function: } K(r) = \frac{{\mathbb{E}\left[ {t\left( {u,r,{\mathbf{X}}} \right)|u \in {\mathbf{X}}} \right]}}{\rho}, \\
        \text{Besag's centered $L$-function: } L(r) = \sqrt {\frac{{K(r)}}{\pi }}  - r
    \end{gathered}
\end{equation}

An estimator for the empirical $K$-function ($\hat K(r)$) is formulated in Equation~\ref{eq:k-funct}, which is the cumulative average number of neighbors within $r$ radius of a typical point, standardised by the intensity and corrected for edge effects.
\begin{equation}\label{eq:k-funct}
  \hat K(r) = \frac{W}{{n(n - 1)}}\sum\limits_{i = 1}^n {\sum\limits_{j = 1,j \ne i}^n {\nu ({d_{ij}} \leqslant r){e_{ij}}(r)} } 
\end{equation} 
where $\nu(.)$ is an indicator function that equals $1$ if the argument is true and otherwise is $0$. Here $n$ is the number of points;  $W$ is the area of the observation window; $r$ is the interaction distance; $d_{ij}$ is the Euclidean distance between $x_i$ and $x_j$; and $e_{ij}$ denotes weights for edge correction \citep{baddeley2015spatial}. The $K$- and $L$-functions can illustrate the non-random spatial arrangement of the points if compared with CSR. They are invariant to the intensity of a point pattern and to missing random points \citep{Ripley1976, Baddeley2000b}, which allows these second-order statistics to be compared when the number of points and observation window vary in the point patterns under consideration. Positive values of the centered $L$-function depict spatial attraction, and negative values describe spatial inhibition in a point pattern. 

\begin{wrapfigure}{t}{0.61\textwidth} 
	\centering
	\includegraphics[width=0.61\textwidth]{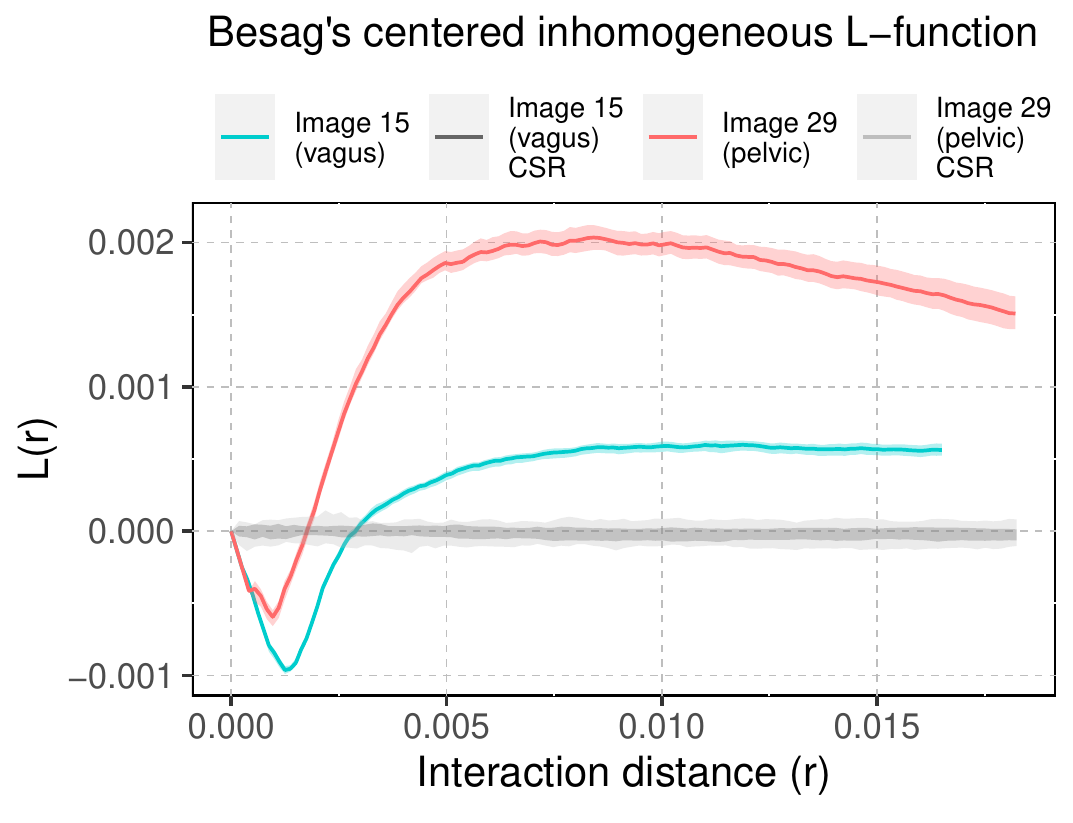}
	\caption{Besag's centered inhomogeneous $L$-function computed for Images 15 and 29. The shaded area around $L(r)$=0 shows the significance bands of complete spatial randomness (CSR). The solid lines illustrate the non-random spatial arrangement of the point patterns compared to CSR. The shaded area around the solid lines shows the boundaries of $95$-percentile confidence interval.}
	\label{fig:Linhom}
\end{wrapfigure}

When dealing with inhomogeneous point patterns, an inhomogeneous $L$-function, based on the inhomogeneous $K$-function ($K_{\textrm{inhom}}(r)$) can be evaluated. The estimator for the inhomogeneous $K$-function ($\hat K_{\textrm{inhom}}(r)$) is formulated in Equation~\ref{eq:kinhom-funct} below.
\begin{equation}\label{eq:kinhom-funct}
\begin{gathered}
    {{\hat K}_{\textrm{inhom}}}\left( r \right) = \frac{1}{{{D^p}W}}\sum\limits_{i=1}^n {\sum\limits_{j=1,j \ne i}^n {\frac{{\nu ({d_{ij}} \leqslant r)}}{{\hat \rho \left( {{x_i}} \right)\hat \rho \left( {{x_j}} \right)}}} } e_{ij}(r), \hfill\\
    {D^p} = {\left( {\frac{1}{W}\sum\limits_{i=1}^n {\frac{1}{{\hat \rho \left( {{x_i}} \right)}}} } \right)^p},\quad p \in \left\{ {1,2} \right\} \hfill \\
\end{gathered}
\end{equation}
where $\hat \rho (u)$ is an estimator of the intensity function $\rho(u)$, obtained using a kernel-smoothed (described later in the paper) intensity estimator \citep{baddeley2015spatial}. Figure~\ref{fig:Linhom} shows the Besag’s centered inhomogeneous L-function computed for Image 15 (vagus) and Image 29 (pelvic), listed in Table~\ref{tab:dataset} and displayed in Figure~\ref{fig:fascicle_segment_pp}. It illustrates spatial inhibition for a small range of interaction distance at the beginning, then spatial attraction for both the samples. The spatial attraction, in other words, the clustering tendency is more apparent in the pelvic sample which agrees with the images shown in Figure~\ref{fig:fascicle_segment_pp}. 

An SPP is \emph{anisotropic} if any of its statistical characteristics change when the point pattern is rotated about any axis in 2-D or 3-D space. The $K$- and $L$-functions can be modified in various ways to estimate anisotropy \citep{ohser_second-order_1981, stoyan_stochastic_2013}. Computing the cumulative distribution of the neighbors within a section of the disc of $r$ radius between two directional preferences $\theta_{1}$ and $\theta_{2}$, instead of the entire disc of $r$ radius, gives the sector $K$- and $L$-functions \citep{baddeley2015spatial}.   

When an SPP exhibits different interactions in different places, it is beneficial to look into the spatial statistics locally by decomposing them into contributions from individual points \citep{baddeley2015spatial}. If the $K$-function estimator ($\hat K(r)$) shown in Equation~\ref{eq:k-funct} is decomposed, the contributions from individual points are referred to as local $K$-functions and can be formulated as follows:
\begin{equation}\label{eq:localk-funct}
    \hat K(r, x_i) = \frac{W}{n-1}\sum\limits_{j=1, j \ne i}^n{\nu ({d_{ij}} \leqslant r){e_{ij}}(r)} , \text{ for } i=1, \dots, n
\end{equation}
The estimator $\hat K(r)$ is simply the average of all the $\hat K(r, x_i)$s  for $i=1, \dots, n$. Arbitrarily, the centered local $L$-functions are formulated as : $\hat L(r, x_i) = (\hat K(r, x_i) / \pi)^{1/2} - r$. This notion of decomposition is applicable for the inhomogeneous and anisotropic $K$- and $L$-functions as well.

The points in an SPP may be of different types (multitype point pattern). The additional information attached to each point in point patterns is called a mark and can hold categorical or continuous-valued, physical or statistical characteristics. The points can carry additional attributes (forming marked point patterns) or be linked to the space of interest (covariates). It is often helpful to apply spatial smoothing to the marks of a point pattern for visualization and various post-processing purposes. The result of the kernel-smoothing (usually with Gaussian kernel) at a location $u$ is a spatially weighted average of the marks attached to the points in the neighborhood of $u$ \citep{baddeley2015spatial}. This is also known as the Nadaraya-Watson smoother \citep{nadaraya_estimating_1964, nadaraya_nonparametric_1989, watson_smooth_1964}.

We intend to utilize the spatial statistics discussed above, specifically spatial intensity, and local inhomogeneous and anisotropic $L$-functions to describe the spatial arrangement of the point patterns constructed from the unmyelinated axons in the nerve cross-sections.

\subsection{Optimal transport framework and Sinkhorn distance}\label{sec:ot}
The \emph{transport problem} 
distributes a certain amount of \emph{mass} from a set of sources to a set of destinations at minimum cost. 
There are two major factors  in a transport problem: the \textit{cost function} and the \textit{transportation plan}. The cost function defines a fixed, non-negative effort required to transport unit mass from a source to a destination. This cost may only depend on the distance between the source and the destination or on other additional factors; in the former case a Euclidean distance matrix between the sources and the destinations is a reasonable representation of effort. 

Once the cost of transportation is represented, the remaining part of the problem involves transporting a non-negative amount of mass between sources and destinations, as described by a transportation plan. Various transportation plans result in different total costs, and the \emph{optimal transport problem} (OT) aims to minimize this cost. The OT problem is \emph{balanced} if the total mass at the sources equals the total mass at the destinations, and \emph{unbalanced} otherwise \citep{Peyre2019}.

Let $r$ and $c$ be two $d$ dimensional vectors representing the amount of mass at the $d$ sources and the $d$ destinations, respectively. The number of sources and destinations could differ, but they can be considered equal without loss of  generality. Let $U(r,c)$ be the set of all non-negative $d\times d$ matrices with row and column summing to $r$ and $c$,  respectively. Any matrix $P \in U(r,c)$ describes a transportation plan that transports the mass in $r$ to   $c$. Given a $d\times d$ cost matrix $M$, the total cost of mapping $r$ to $c$ using the transportation plan $P$ is $\sum_{i,j}P_{ij}M_{ij}$. Thus the OT problem between $r$ and $c$ given cost $M$ can be formulated 
by Equation~\ref{eq:ot}, where $D_{M}(r,c)$ is the optimal transport distance: 
\begin{equation}\label{eq:ot}
    \begin{gathered}
        D_{M}(r,c) = \min_{P \in U(r,c)}\sum_{i,j}P_{ij}M_{ij},\\
     \text{subject to}\quad \sum_{j}P_{ij} = r_{i},\quad \sum_{i}P_{ij} = c_{j},\quad P_{ij} \ge 0,\quad  \forall i,j \leq d. 
    \end{gathered}
\end{equation}
The masses in $r$ and $c$ could  be normalized to sum to one, and then both $r$ and $c$ can be interpreted as probability distributions.

For $D_{M}(r,c)$ to be a metric, the cost matrix $M$ has to be a metric matrix \citep{villani_wasserstein_2009, avis_extreme_1980, brickell_metric_2008} satisfying the conditions shown in Equation~\ref{eq:metric_space}.
\begin{equation}\label{eq:metric_space}
    \begin{gathered}
        \text{Non-negativity: } M_{ij} \ge 0,  \\
        \text{Identity: } M_{ii}=0,  \\
        \text{Symmetry: } M_{ij}=M_{ji}, \\
        \text{Triangle inequality: } M_{ij} \leq M_{ik} + M_{kj}, \quad \forall i,j,k \leq d.
    \end{gathered}
\end{equation}

An OT problem is a convex optimization problem that can be solved using various approaches \citep{ahuja_network_1993, orlin_faster_1993} and for a general cost matrix the computational cost scales as  $\mathcal{O}(d^3log(d))$  \citep{pele_fast_2009}, which prevents scaling the solution to large problem sizes.  
Earlier approximate solutions obtained by putting constraints on the cost matrix could result in a loss of applicability and performance \citep{grauman_fast_2004}. A later approximation to the original OT problem using an entropic regularization scheme was proposed by \cite{cuturi_sinkhorn_2013} to reduce the computational complexity. The scheme employs the Sinkhorn-Knopp matrix scaling algorithm \citep{sinkhorn_concerning_1967, knight_sinkhornknopp_2008},
and hence the name \emph{Sinkhorn distance} for its objective function.

\subsubsection{Sinkhorn distance}\label{sec:sinkhorn}
A straightforward way of thinking about a transportation plan is by noticing that if a source contains more mass, it should originate more, and if a destination requires more mass, it should receive proportionally more. Such a transportation plan is represented by $rc^{T}$, and the optimal plan $P$ should be somewhere around the distribution $rc^{T}$. Simply speaking, the idea of the entropic regularization scheme by \cite{cuturi_sinkhorn_2013} is to choose $P$ from a smaller set  near $rc^{T}$, instead of the entire set $U(r,c)$.

To capture these ideas, \cite{cuturi_sinkhorn_2013} imposes an additional constraint of Kullback-Leibler (KL) divergence on the OT formulation, as shown in Equation~\ref{eq:sinkhorn}, and computes the Sinkhorn distance $D_{M, \alpha}^{*}(r,c)$. This constraint introduces a set $U_{\alpha}(r, c) \subset U(r, c)$ from which  an optimal transportation plan $P$ is selected. The KL divergence distance between $P$ and $rc^{T}$ is set to be  smaller than a predefined parameter $\alpha$. In other words, $P$ should belong to a distribution near $rc^{T}$. 

\begin{equation}\label{eq:sinkhorn}
    \begin{gathered}
        D_{M, \alpha}^{*}(r,c) = \min_{P \in U_{\alpha}(r,c)}\sum_{i,j}P_{ij}M_{ij},\\
        \text{subject to}\quad \textbf{KL}(P|rc^{T}) \leq \alpha, \quad \sum_{j}P_{ij} = r_{i}, \quad \sum_{i}P_{ij} = c_{j}, \quad  \forall i,j \leq d. 
    \end{gathered}
\end{equation}

The entropy ($h$) of the transportation plan ($P$) and the mass vectors ($r$ and $c$) are given in Equation~\ref{eq:entropy}:
\begin{equation}\label{eq:entropy}
    \begin{gathered}
         h(P) = -\sum_{ij}P_{ij}\log P_{ij}, \\
         h(r) = -\sum_{i}r_{i}\log r_{i}, \quad h(c) = -\sum_{j}c_{j}\log c_{j}. 
    \end{gathered}
\end{equation}
We proceed to express the KL divergence constraint in terms of the entropy: 
\begin{equation}
    \begin{aligned}
        \textbf{KL}(P|rc^{T}) & = \sum_{ij}P_{ij}\log\frac{P_{ij}}{r_{i}c_{j}}\\
        & = \sum_{ij}P_{ij}\log P_{ij} - \sum_{ij}P_{ij}\log r_{i} - \sum_{ij}P_{ij}\log c_{j}\\
        & = \sum_{ij}P_{ij}\log P_{ij} - \sum_{i}r_{i}\log r_{i} - \sum_{j}c_{j}\log c_{j} \quad [\because \sum_{j}P_{ij} = r_{i}, \sum_{i}P_{ij} = c_{j}] \\
        & = -h(P) + h(r) + h(c) \leq \alpha. 
    \end{aligned}
\end{equation}
Thus the new constraint states that the entropy of $P$ should be large enough to satisfy 
$$h(P) \geq h(r) + h(c) - \alpha, $$ 
which constrains $P$ to be chosen from the Kullback-Leibler ball of level $\alpha$ centered about $rc^T$ (see Figure 1 in \citep{cuturi_sinkhorn_2013}). 

This interpretation makes the OT problem non-convex, and an alternative formulation of Sinkhorn distance is required for ease of optimization. For every pair $(r,c)$, each $\alpha$ corresponds to a  Lagrange multiplier $\lambda \in [0, \infty)$ such that $D_{M, \alpha}^{*}(r,c) = D_{M}^{\lambda}(r,c)$. 
The distance $D_{M}^{\lambda}$, shown in Equation~\ref{eq:dual-sinkhorn}, is called the dual-Sinkhorn divergence by \cite{cuturi_sinkhorn_2013}.

\begin{equation}\label{eq:dual-sinkhorn}
    \begin{gathered}
        D_{M}^{\lambda}(r,c) = \sum_{i,j}P^{\lambda}_{ij}M_{ij}, \quad \text{where } P^{\lambda}= \argmin{P \in U(r,c)}\sum_{i,j}P_{ij}M_{ij} - \lambda h(P),\\
        \text{subject to}\quad \sum_{j}P_{ij} = r_{i}, \quad  \sum_{i}P_{ij} = c_{j},\quad \forall i,j \leq d. 
    \end{gathered}
\end{equation}

By introducing two dual variables  $\phi$ and $\psi$ for each of the two equality constraints of Equation~\ref{eq:dual-sinkhorn}, the Lagrangian of the objective function can be written as Equation~\ref{eq:lagrange}.

\begin{equation}\label{eq:lagrange}
    \mathcal{L}(P, \phi, \psi) =  \sum_{i,j}P_{ij}M_{ij} - \lambda h(P) + \sum_{i}\phi_{i}(\sum_{j}P_{ij} - r_{i}) + \sum_{j}\psi_{j}(\sum_{i}P_{ij} - c_{j}). 
\end{equation}

The  derivative of the Lagrangian objective function  with respect to $P_{ij}$, for any pair $(i,j)$, can be set to zero to obtain an extremum; the second derivative of the Lagrangian,  ($\frac{\lambda}{P_{ij}}$), is positive since 
both the numerator and the denominator are positive, and thus we have obtained a minimizer of the Lagrangian. 
\begin{equation}\label{eq:lagrange_derivative}
    \begin{aligned}
        \frac{\partial \mathcal{L}}{\partial P_{ij}} &= M_{ij} + \lambda + \lambda \log P_{ij} + \phi_{i} + \psi_{j} = 0.  \\
        \implies \ P_{ij} &= e^{-\frac{\phi_{i}}{\lambda}-\frac{1}{2}}. e^{-\frac{M_{ij}}{\lambda}}. e^{-\frac{\psi_{j}}{\lambda}-\frac{1}{2}}\\
        &\equiv u_{i} K_{ij} v_{j} \quad [u_{i} = e^{-\frac{\phi_{i}}{\lambda}-\frac{1}{2}}, v_{j}=e^{-\frac{\psi_{j}}{\lambda}-\frac{1}{2}}, K = e^{-\frac{M}{\lambda}}] 
    \end{aligned}
\end{equation}

Given $K$, $r$ and $c$, the Sinkhorn-Knopp matrix scaling algorithm converges to a solution
$P^{\lambda}$ of the following form:
\begin{equation}
    \exists u, v: P^{\lambda} = \textbf{diag}(u) K \textbf{diag}(v).
\end{equation}
$P^{\lambda}$ should have the correct row and column sums, as shown in Equation~\ref{eq:dual-sinkhorn}. We deduce the update rule for the Sinkhorn-Knopp algorithms from those constraints in the following manner:
\begin{align*}
    &\sum_{j}P_{ij}^{\lambda} = r_{i}, & & \sum_{i}P_{ij}^{\lambda} = c_{j}\\
    \implies & \sum_{j}u_{i} K_{ij} v_{j} = r_{i} \quad [\text{Equation}~\ref{eq:lagrange_derivative}] & \implies & \sum_{i}u_{i} K_{ij} v_{j} = c_{j} \quad [\text{Equation}~\ref{eq:lagrange_derivative}]\\
    \implies & u_{i} \sum_{j}K_{ij} v_{j} = r_{i} & \implies & v_{j} \sum_{i}u_{i} K_{ij} = c_{j}\\
    \implies & u_{i} = r_{i} / \sum_{j}K_{ij} v_{j} & \implies & v_{j}  = c_{j} / \sum_{i}u_{i} K_{ij}
\end{align*}

Thus the update rule for the Sinkhorn-Knopp algorithm can be written as Equation~\ref{eq:update_rule}, where $v$ can be initialized randomly.
\begin{equation}\label{eq:update_rule}
    \begin{gathered} 
        u = r ./ (Kv),\\
        v = c ./ (K^{T}u). 
    \end{gathered}
\end{equation}

\cite{cuturi_sinkhorn_2013} observes that the number of iterations in the Sinkhorn-Knopp algorithm is bounded independent of $d$. Thus, the cost of computing $D_{M}^{\lambda}$ is $\mathcal{O}(d^{2})$, which is an improvement over $\mathcal{O}(d^3log(d))$. \cite{cuturi_sinkhorn_2013} describes an approach to compute the Sinkhorn distance ($D_{M, \alpha}^{*}(r,c)$) through the dual-Sinkhorn divergence ($D_{M}^{\lambda}(r,c)$), and also reports that the dual-Sinkhorn divergence does not perform worse than the classic optimal transport distances. Therefore, we use the dual-Sinkhorn divergence to measure the distance between the spatial statistics of the point patterns in our experiments and refer to as the Sinkhorn distance. We utilize the R packages \textit{T4transport} \citep{T4transport} and \textit{Barycenter} \citep{Barycenter} for computing the dual-Sinkhorn divergences, and \textit{spatstat} \citep{baddeley2015spatial} for spatial point pattern analysis.

\section{Experiments and results}\label{sec:experiments}

We represent the unmyelinated axonal arrangements in the vagus and pelvic nerve cross-sections as spatial point patterns. We intend to quantify (using the Sinkhorn distance) similarities between the point patterns in terms of the following spatial features: 
\begin{enumerate}
    \item spatial intensity,
    \item local inhomogeneous $L$-function,
    \item local inhomogeneous anisotropic $L$-function with
    \begin{enumerate}
        \item horizontal and
        \item vertical sectors.
    \end{enumerate}
\end{enumerate}
For the horizontal and vertical cases, we choose sectors forming 15\degree  segment around the horizontal (0\degree) and vertical (90\degree) axes, respectively. We attach the above-mentioned spatial features to the point patterns as marks (described in Section~\ref{sec:spp}). We compute the Sinkhorn distance between every pair of point patterns, for the four spatial features, in two different manners: 
\begin{enumerate}
    \item using the spatial point patterns directly (in Section~\ref{subsec:exp1}) and
    \item using the map of the spatial features constructed by kernel-smoothing (in Section~\ref{subsec:exp2}).
\end{enumerate}
The Sinkhorn distance between every pair of nerve cross-sections is then used to construct a symmetric Sinkhorn distance matrix and visualized in an embedded space via multi-dimensional scaling. In the following three subsections, we describe a few preprocessing and parameter selection tasks required for configuring the spatial features, before going into the experimental details.

\subsection{Interaction distance configuration}\label{subsec:intr_dist}
A critical issue regarding the computation of local inhomogeneous $L$-functions is determining the interaction distance ($r$), described in Section~\ref{sec:spp}. The point patterns constructed from the nerve cross-sections differ in size, as do their interaction ranges. The preferred choice for the interaction distance is the one that can reasonably separate the spatial features of the point patterns. We compute a range of interactions that are common for all the point patterns and configure the interaction distance using two approaches: (a) based on the standard deviations of the inhomogeneous $L$-function of all the point patterns and (b) based on the F-ratio (analysis of variance) of the inhomogeneous $L$-function of the point patterns grouped as vagus vs. pelvic, within the expected range.

\subsection{Translation and rotation normalization}\label{subsec:trans_rot_norm}
The optimal transport distance is not invariant under translations and rotations \citep{Wang2013c}. This is critical in the case of analyzing the point patterns because spatial inhomogeneity and anisotropy depend largely on the placement and orientation of the point patterns. To provide translation invariance, we scale the point patterns maintaining proportionality, align the center of mass to the origin, and apply the necessary $0$-padding around the biological structures. 

Ensuring rotation invariance is non-trivial. In an ideal setting, the information regarding the orientation of the biological structures would be available directly to the analyst. Unfortunately, the experimental and instrumental setting may not always allow the orientation of the samples to be maintained during the specimen preparation and the imaging process. Therefore, we implemented a post-hoc minimization process as a workaround. While computing the Sinkhorn distance between the spatial intensities of a pair of point patterns, we keep the orientation of one of them unchanged and rotate the other one about the origin by multiple $\theta$ values ($\theta=$45\degree in our experiments). We compute the Sinkhorn distance for all possible values of $\theta$ and keep the orientation that provides the smallest Sinkhorn distance result. We use this identified orientation for the computation of other spatial features. This method provides a reproducible procedure in the absence of known anatomical orientation data. 

\subsection{The entropic regularization parameter}\label{subsec:lambda}
We refer to the coefficient of the entropy of the transportation plan ($h(P)$) in the dual-Sinkhorn divergence formulation shown in Equation~\ref{eq:dual-sinkhorn}, $\lambda$, as the entropic regularization parameter. As $\lambda\rightarrow 0$, Sinkhorn distance approaches the optimal transport distance (Wasserstein distance, provided that the cost is Euclidean distance). As $\lambda$ increases, the computation results in different approximations of the optimal transport distance, i.e., the Sinkhorn distances. Tuning the appropriate entropic regularization parameter is an important task. We can consider two scenarios: (a) selecting an entropic regularization parameter that provides better separation between analyzed instances and (b) selecting an entropic regularization parameter that makes the Sinkhorn distance a more accurate approximation of the exact optimal transport distance. Thus, parameter tuning is a trade-off between favoring the utility of the method (and lower computational cost) and the accuracy of the approximation. We experimented with $\lambda$ = \{0.01, 0.05, 0.1, 0.5, 1.0, 2.0, 5.0\}, and report the the results for $\lambda$=0.01.

\subsection{Sinkhorn distance between spatial point patterns}\label{subsec:exp1}
In this section, we compute Sinkhorn distance between the spatial point patterns (directly) of the nerve cross-sections, for the four spatial features mentioned earlier. Let $S_{1}$ and $S_{2}$ be two spatial point patterns, with $n_{1}$ and $n_{2}$ number of points respectively. Considering an optimal transport problem between $S_{1}$ and $S_{2}$, we assume that each point in $S_{1}$ contains $\frac{1}{n_{1}}$ amount of mass ($r$), therefore the total mass $=\frac{1}{n_{1}}\times n_{1} = 1$. Similarly, each point in $S_{2}$ requires $\frac{1}{n_{2}}$ amount of mass ($c$), so the total mass $=\frac{1}{n_{2}}\times n_{2} = 1$. Thus the problem is to transport the mass from $S_{1}$ to $S_{2}$ (balanced). The inputs for the computations are the spatial locations of the points in $S_{1}$ and $S_{2}$. Therefore, we can compute the Euclidean distance matrix between them to be the cost matrix $M$, of dimension $n_{1}\times n_{2}$. Here, the cost matrix $M$ captures the spatial intensity of the point patterns. Further, we compute the transportation plan $P$, of dimension $n_{1}\times n_{2}$, using the formulation described in Section~\ref{sec:sinkhorn} and obtain the Sinkhorn distance, which provides a measure of similarity of the spatial intensity between $S_{1}$ and $S_{2}$.

In the cases of the three other spatial features, the cost matrix ($M$) remains the same (the spatial location of the points are unchanged), but the amount of mass produced ($r$) or required ($c$) at each point changes. The local inhomogeneous $L$-function attaches a numerical value to each point in a point pattern that captures its local spatial interaction within a certain interaction distance. Instead of a uniform mass amount, we may assume that each point in $S_{1}$ and $S_{2}$ is assigned an amount of mass that equals its local inhomogeneous $L$-function value. We normalize the values assigned to each point pattern to sum to one. Then we can compute the transportation plan $P$ in the same manner as described above. The resultant Sinkhorn distance gives us a measure of similarity of the local spatial interaction between $S_{1}$ and $S_{2}$. The Sinkhorn distances for the anisotropic spatial features, the local inhomogeneous anisotropic $L$-function with horizontal and vertical sectors, are also computed similarly.

\begin{figure}
    \centering
    \includegraphics[width=\textwidth]{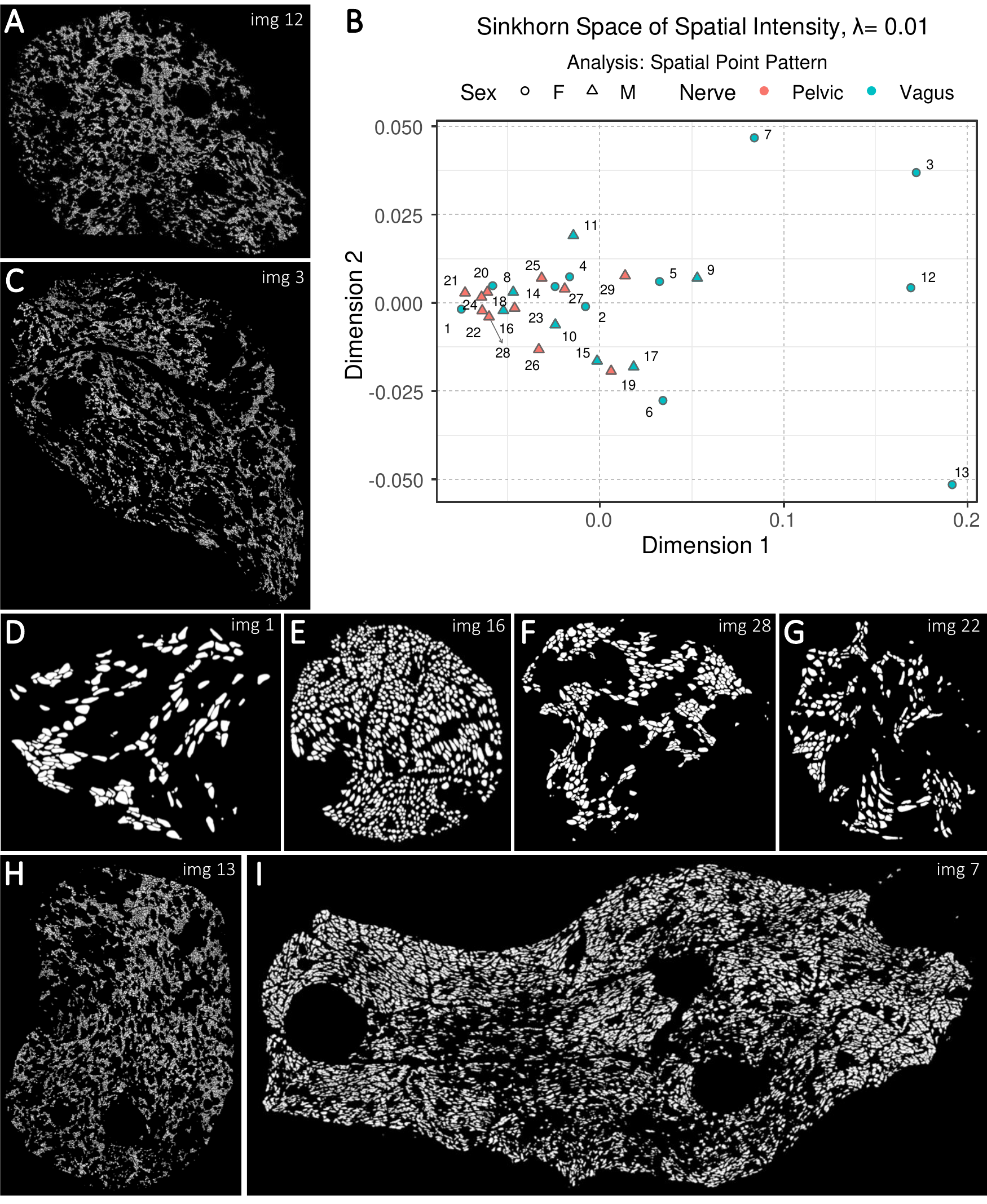}
    \caption{(A, C-I) A set of images of the segmented unmyelinated axons in the nerve cross-sections, labeled with the Image ID. (B) An embedding of the spatial intensity of the spatial point patterns in the Sinkhorn space for entropic regularization parameter $\lambda = 0.01$. The vagus and the pelvic samples are shown in cyan and orange (circles for female (F) and triangles for male (M)), respectively, and labeled with the Image ID listed in Table~\ref{tab:dataset}.}
    \label{fig:t4_si_1}
\end{figure}

We have 29 nerve cross-sections in the dataset and once we compute the Sinkhorn distance between every pair, we can construct Sinkhorn distance matrices of dimension 29$\times$ 29, for each of the four spatial features. We use muti-dimensional scaling to embed the Sinkhorn distance matrices in 2-D to illustrate the computed Sinkhorn distance between the spatial features and interpret the notion of similarity (or dissimilarity) between the nerve cross-sections. We denote the new embedded 2-D space as the \textit{Sinkhorn space}.

Figure~\ref{fig:t4_si_1}B shows an embedding of the spatial intensity of the spatial point patterns (used directly) in the Sinkhorn space for entropic regularization parameter $\lambda $=0.01. The vagus and the pelvic samples are shown in cyan and orange, respectively, and labeled with the Image ID listed in Table~\ref{tab:dataset}. Figure~\ref{fig:t4_si_1}H shows Image 13 (vagus), which is positioned far from the other samples in the Sinkhorn space. It is the largest sample in our dataset regarding image size and the number of segmented unmyelinated axons. It is also the only sample from a left cervical trunk. Figure~\ref{fig:t4_si_1}A and Figure~\ref{fig:t4_si_1}C display Image 12 and Image 3 (both vagus) respectively. They are collected from the right cervical trunks. Figure~\ref{fig:t4_si_1}I shows Image 7 (vagus) from abdominal vagus posterior trunk. Considering the spatial intensity, these four vagus samples are embedded far apart and are visually different. The rest of the samples are positioned in proximity, yet we can see a rightward tendency in the vagus samples than the pelvic ones. Figure~\ref{fig:t4_si_1}(D, E) show Image 1 and Image 16 (both vagus) respectively. They look spatially different from the vagus samples discussed so far and are embedded at the leftmost part of the Sinkhorn space, far from those samples. However, they have a similar spatial organization as Image 28 and Image 22 (both pelvic) shown in Figure~\ref{fig:t4_si_1}(F, G) and are embedded closer in the Sinkhorn space. 

Although one can intuitively understand the global differences in spatial intensity of the point patterns by looking at the images of segmented unmyelinated axons in the nerve cross-sections, our approach can quantify and visualize the differences with the Sinkhorn distance between every pair of samples, resulting in a map of patterns. For instance, in Figure~\ref{fig:t4_si_1}B, the Sinkhorn distances between the spatial intensity of Image 1, and Image 3 and Image 12 are 0.257 and 0.254 respectively, whereas the distance between Image 3 and Image 12 is 0.107. Again, Image 16 and Image 28 have respectively distances 0.054 and 0.048 from Image 1.

\begin{figure}
    \centering
    \includegraphics[width=\textwidth]{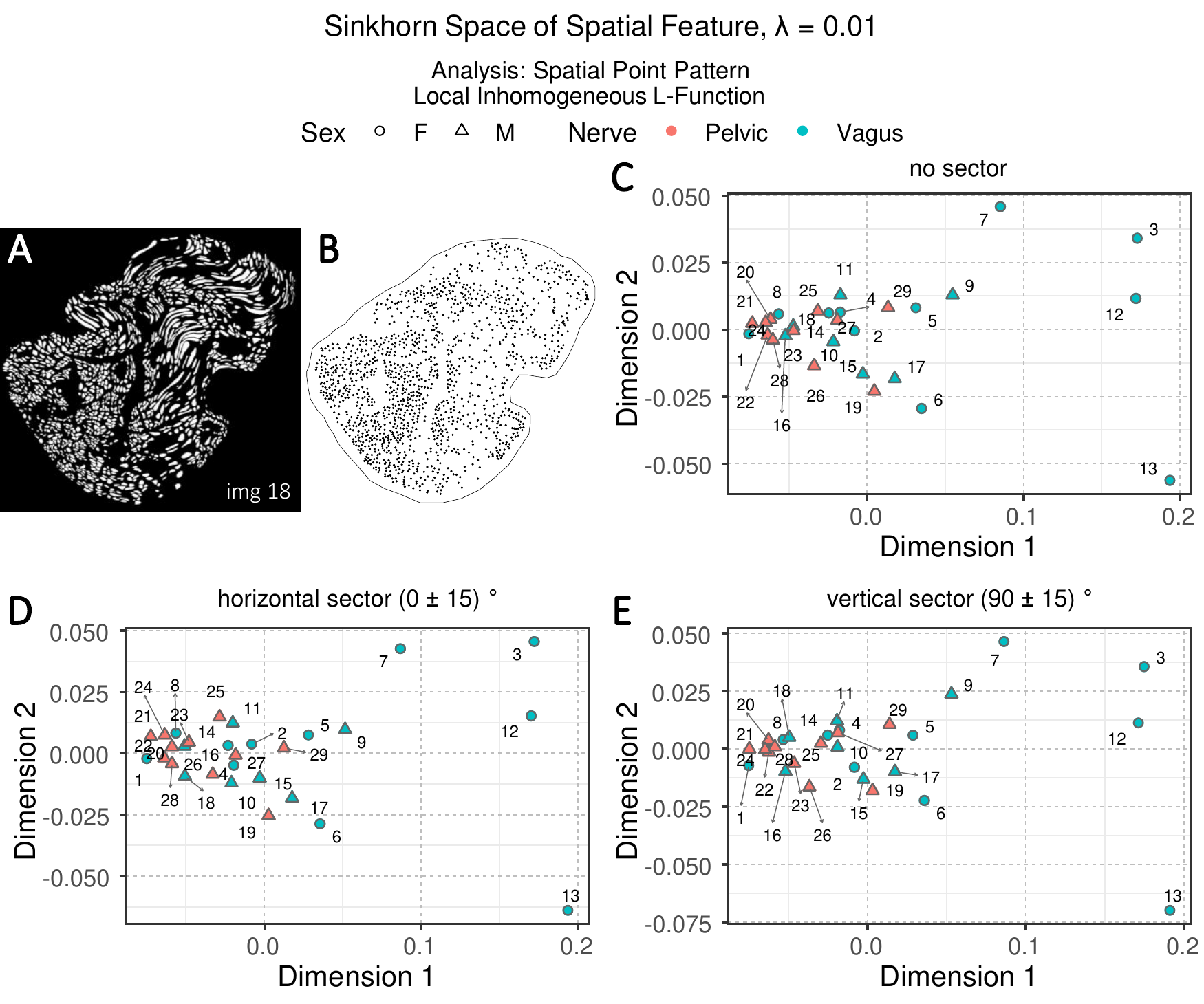}
    \caption{(A, B) The segmented unmyelinated axons and the spatial point pattern of Image 18 (vagus) listed in Table~\ref{tab:dataset}. (C-E) The embeddings of the local inhomogeneous and anisotropic $L$-functions (no sector, horizontal sector, and vertical sector) of the spatial point patterns in the Sinkhorn space for entropic regularization parameter $\lambda = 0.01$. The vagus and the pelvic samples are shown in cyan and orange (circles for female (F) and triangles for male (M)), respectively, and labeled with the Image ID listed in Table~\ref{tab:dataset}.}
    \label{fig:t4_sf_1}
\end{figure}

Interpreting spatial statistics, such as local inhomogeneous and anisotropic $L$-functions, can be more challenging than understanding raw spatial intensity. Figure~\ref{fig:t4_sf_1}(C-E) show three embeddings of the spatial features in the Sinkhorn space for $\lambda$=0.01: the local inhomogeneous $L$-function, the local inhomogeneous $L$-function with horizontal sector and vertical sector, respectively. With a few exceptions, the overall landscape in the embeddings is similar to the one for the spatial intensity shown in Figure~\ref{fig:t4_si_1}B. Figure~\ref{fig:t4_sf_1}A, showing the segmented unmyelinated axons for Image 18 (vagus) contains several elongated axons. The elongated axons make the spatial arrangement of centroids in the corresponding point pattern (see Figure~\ref{fig:t4_sf_1}B) quite sparse and direction-oriented (anisotropic) in certain regions. The different positioning of Image 18 in the embeddings can reflect these characteristics. 

\subsection{Sinkhorn distance between maps of spatial features}\label{subsec:exp2}
Here, we compute the Sinkhorn distance between every pair of point pattern using the map of the spatial features constructed by kernel-smoothing. When we consider spatial point patterns directly to compute the Sinkhorn distance between nerve cross-sections (in Section~\ref{subsec:exp1}), the mass (corresponding to the spatial intensity or any other spatial feature attached as marks) to be transported is concentrated at the exact location of the points. As we apply kernel-smoothing to the point pattern, the concentrated mass at any point diffuses into its neighborhood. This phenomenon can help capture the essence of locality in the kernel-smoothed maps while computing Sinkhorn distances. In addition, kernel-smoothing can reduce the influence of the differing number of points in the point patterns on the Sinkhorn distance, by constructing maps of the same dimensions. Also, transportation-based metrics are well-suited for quantifying differences between bitmap images in which pixel values can be interpreted as transportable mass without strict geometric constraints \citep{rubner_earth_2000, haker_optimal_2004, chefd2007intensity, grauman_fast_2004, wang_optimal_2011, wang_linear_2013}.

\begin{figure}
    \centering
    \includegraphics[width=\textwidth]{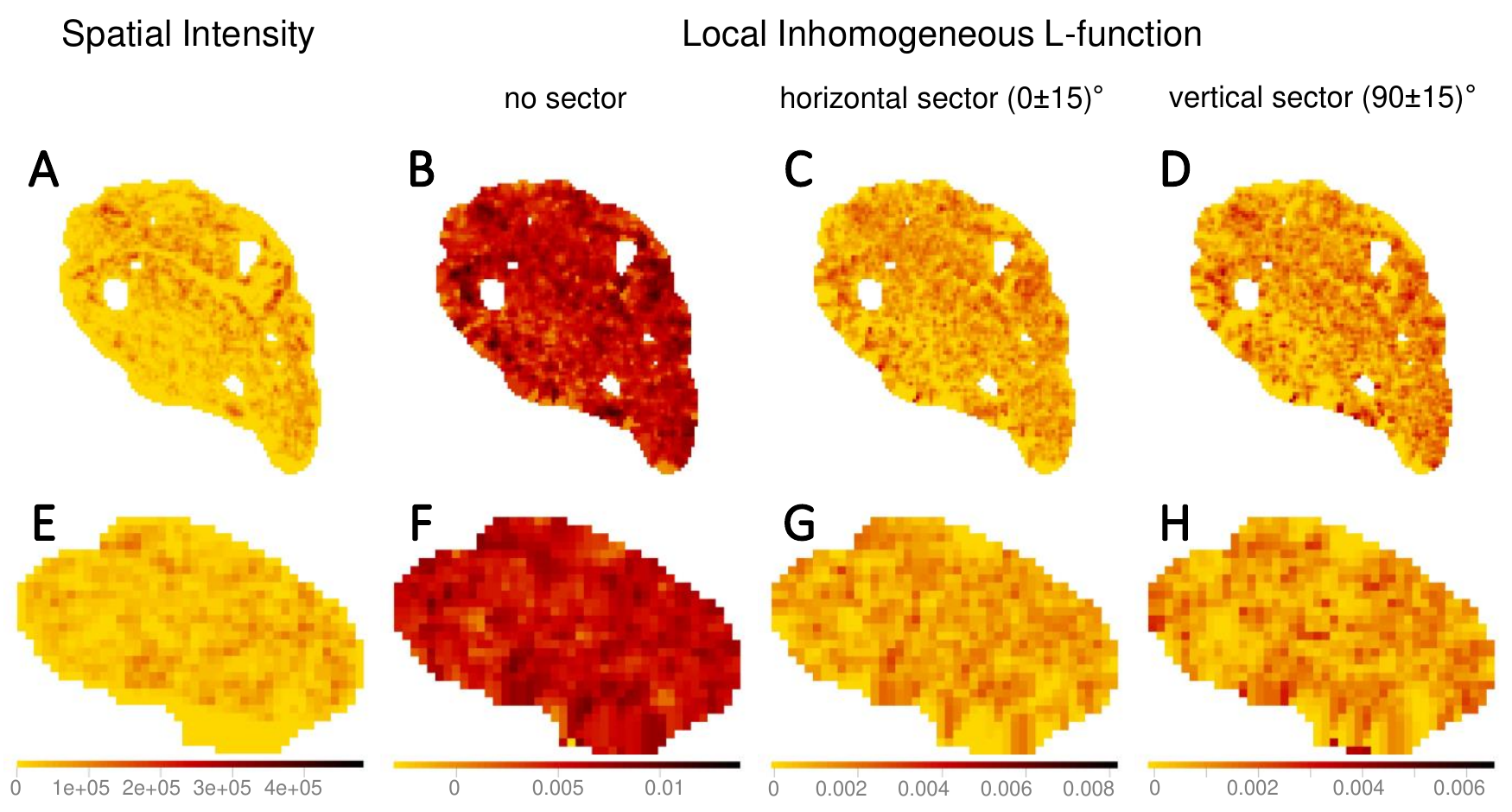}
    \caption{Visualizing the kernel-smoothed spatial features of Image 3 (vagus) and Image 29 (pelvic) listed in Table~\ref{tab:dataset}. (A, E) Spatial intensity. (B, F) Local inhomogeneous $L$-function. (C, G) Local inhomogeneous $L$-function with the horizontal sector. (D, H) Local inhomogeneous $L$-function with the vertical sector. The scale bars show the range of values for each spatial feature separately (column-wise). The kernel-smoothed bitmaps were downsampled for reasonable runtime and memory requirements.}
    \label{fig:heatmap}
\end{figure}

The kernel-smoothed intensity and marks of a point pattern can be represented as bitmaps, where the pixel values represent kernel-smoothed intensity or other derived quantities from the marks, such as the local inhomogeneous and anisotropic $K$- and $L$-functions. The kernel-smoothed spatial features of Images 3 (vagus) and 29 (pelvic) listed in Table~\ref{tab:dataset} are shown in Figure~\ref{fig:heatmap}. The bitmaps for the local inhomogeneous $L$-function, shown in Figure~\ref{fig:heatmap}(B, F), have higher values of the spatial feature compared to their anisotropic counterparts shown in  Figure~\ref{fig:heatmap}(C, D, G, H) and slight shifts in values at certain locations are observed between the bitmaps of the horizontal and vertical sectors. Quantifying similarities between the kernel-smoothed bitmaps can be performed just like quantifying similarities between the spatial features of the original point patterns, using Sinkhorn distance.

Let $I_{1}$ and $I_{2}$ be the centered ($0$-padded as necessary) kernel-smoothed maps (same dimension) of the spatial intensity of the point patterns $S_{1}$ and $S_{2}$, respectively. The pixel values in $I_{1}$ and $I_{2}$ are normalized to sum to one, and the value at each pixel is considered the amount of mass contained ($r$) or required ($c$) at that pixel. The location of the pixels is not known beforehand, so we construct a unique grid $[0,1]^{2}$ over which the pixel locations of $I_{1}$ and $I_{2}$ are defined. The cost matrix $M$ is the Euclidean distance matrix computed from the $[0,1]^{2}$ grid. The transportation plan $P$ and the Sinkhorn distance between $I_{1}$ and $I_{2}$ are computed in the previously described manner. The Sinkhorn distances between the maps representing the other three spatial features are also calculated in the same fashion. Constructing the Sinkhorn distance matrix and visualizing embeddings in the Sinkhorn space are also done in the same way as described in Section~\ref{subsec:exp1}.

\begin{figure}[tbh!]
    \centering
    \includegraphics[width=\textwidth]{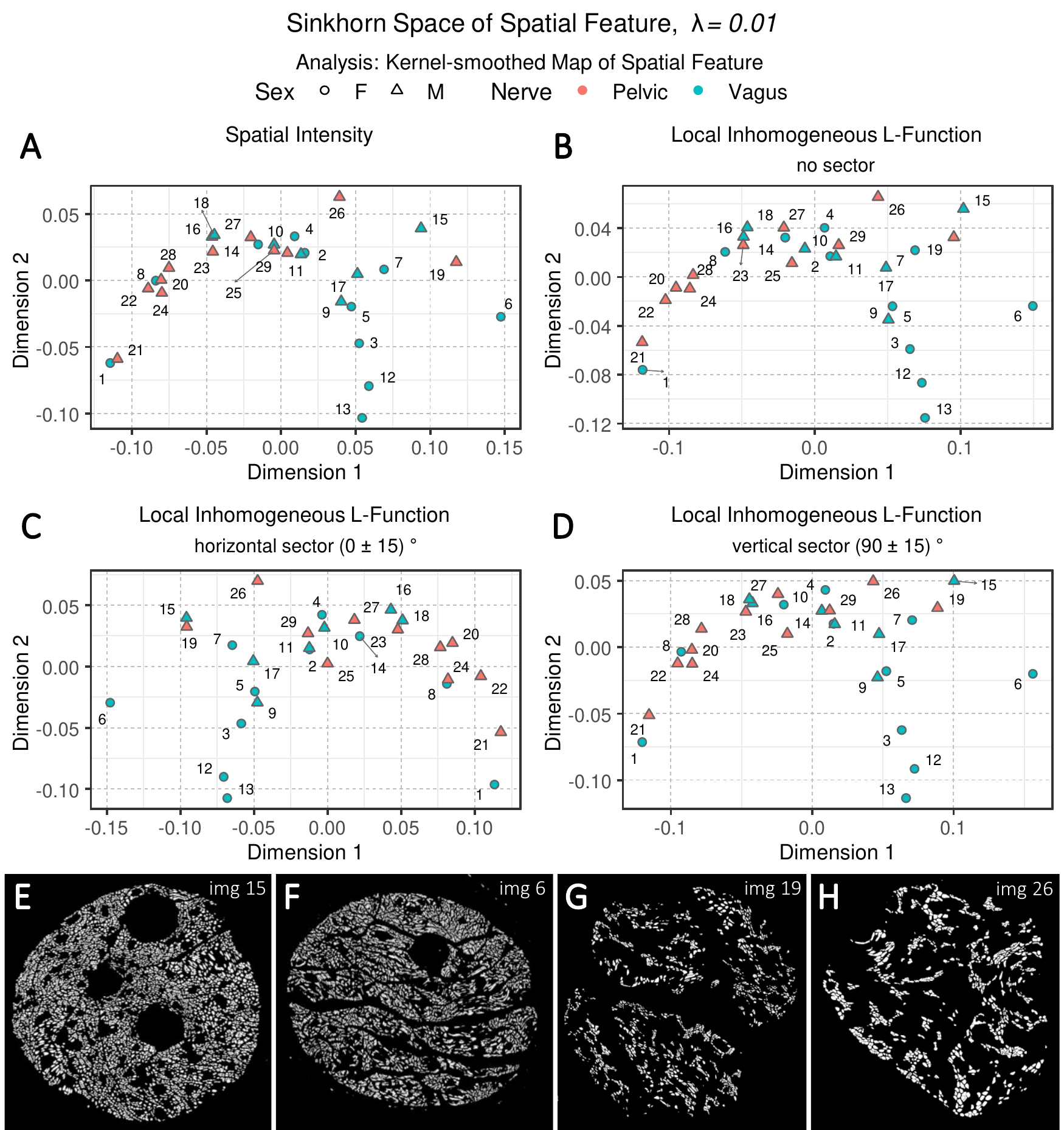}
    \caption{(A-D) The embeddings of the kernel-smoothed spatial intensity and the local inhomogeneous and anisotropic $L$-functions (no sector, horizontal sector, and vertical sector) of the spatial point patterns in the Sinkhorn space for entropic regularization parameter $\lambda = 0.01$. The vagus and the pelvic samples are shown in cyan and orange (circles for female (F) and triangles for male (M)), respectively, and labeled with the Image ID listed in Table~\ref{tab:dataset}. (E-H) The segmented unmyelinated axons of Image 15 and 6 (vagus) and Image 19 and 26 (pelvic) listed in Table~\ref{tab:dataset}. }
    \label{fig:bary_sf_1}
\end{figure}

Figure~\ref{fig:bary_sf_1}A shows an embedding of the kernel-smoothed maps of the spatial intensity of the point patterns in the Sinkhorn space of $\lambda$=0.01. The vagus and the pelvic samples are shown in cyan and orange, respectively, and labeled with the Image ID listed in Table~\ref{tab:dataset}. The vagus samples 3, 7, 12, and 13 are embedded at a distance from the rest of the samples, and this trend was also observed in Figure~\ref{fig:t4_si_1}B, when we experimented on point patterns directly. However, vagus sample 6 and 15, and pelvic samples 19 and 26 (see Figure~\ref{fig:bary_sf_1}(E-H)), that were positioned close to the rest of the samples in Figure~\ref{fig:t4_si_1}B, are embedded far apart in the right-most region of the embedding in Figure~\ref{fig:bary_sf_1}A. Therefore, some behavior in the spatial intensity that were not noticed while dealing with point patterns, have surfaced when maps of the spatial features are used. The embeddings of kernel-smoothed local inhomogeneous and anisotropic $L$-function are illustrated in Figure~\ref{fig:bary_sf_1}(B-D), portraying a similar trend overall, where the visually and spatially comparable vagus and pelvic samples are positioned in proximity, the rest are far apart, and the vagus samples are more spread out.


\subsection{Insights regarding spatial architecture}
\begin{figure}[t]
    \centering
    \includegraphics[width=\textwidth]{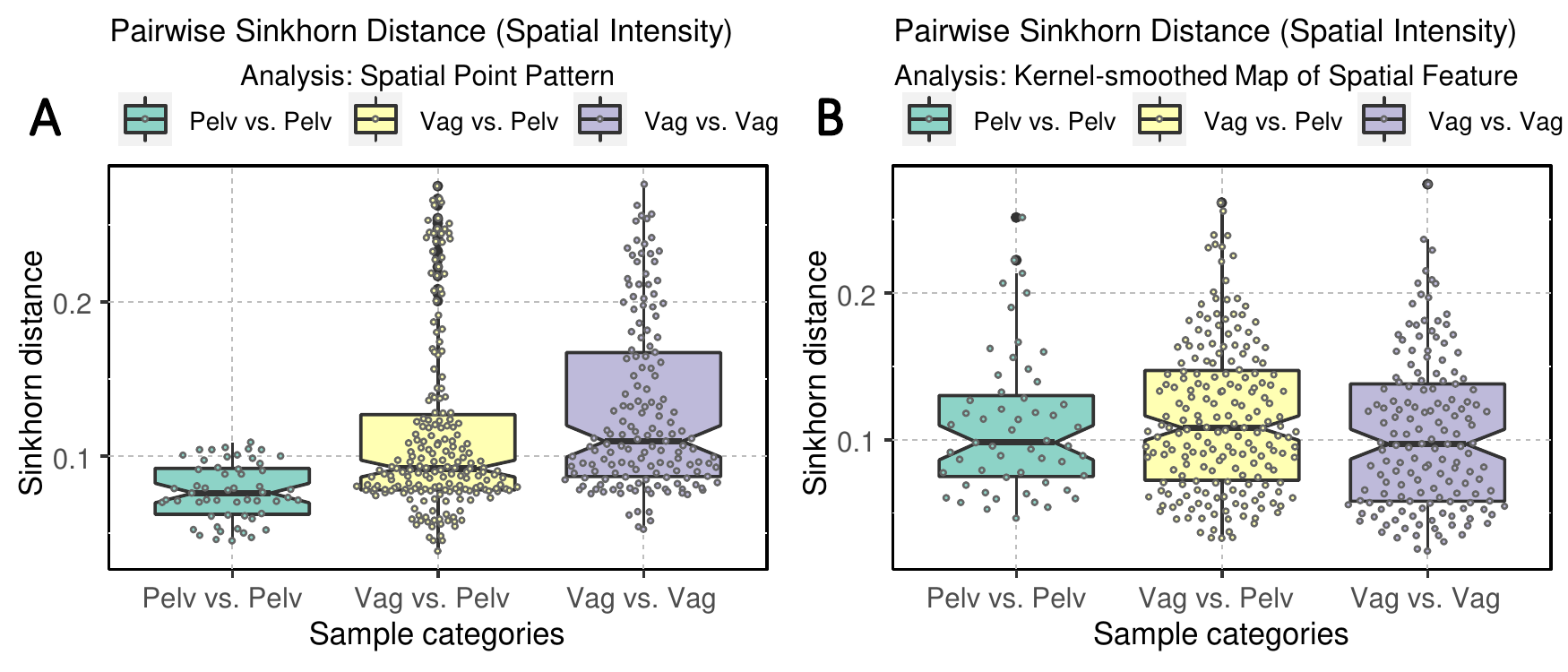}
    \caption{Boxplots displaying the ranges of Sinkhorn distance between the spatial intensity of the nerve cross-sections - Pelv: pelvic and Vag: vagus. (A) Analysis of the spatial point patterns directly. (B) Analysis of the kernel-smoothed maps of the spatial features. The points show the individual Sinkhorn distances, revealing the hidden distribution.}
    \label{fig:t4_Bary_si_stat_2}
\end{figure}

\begin{figure}[t]
    \centering
    \includegraphics[width=\textwidth]{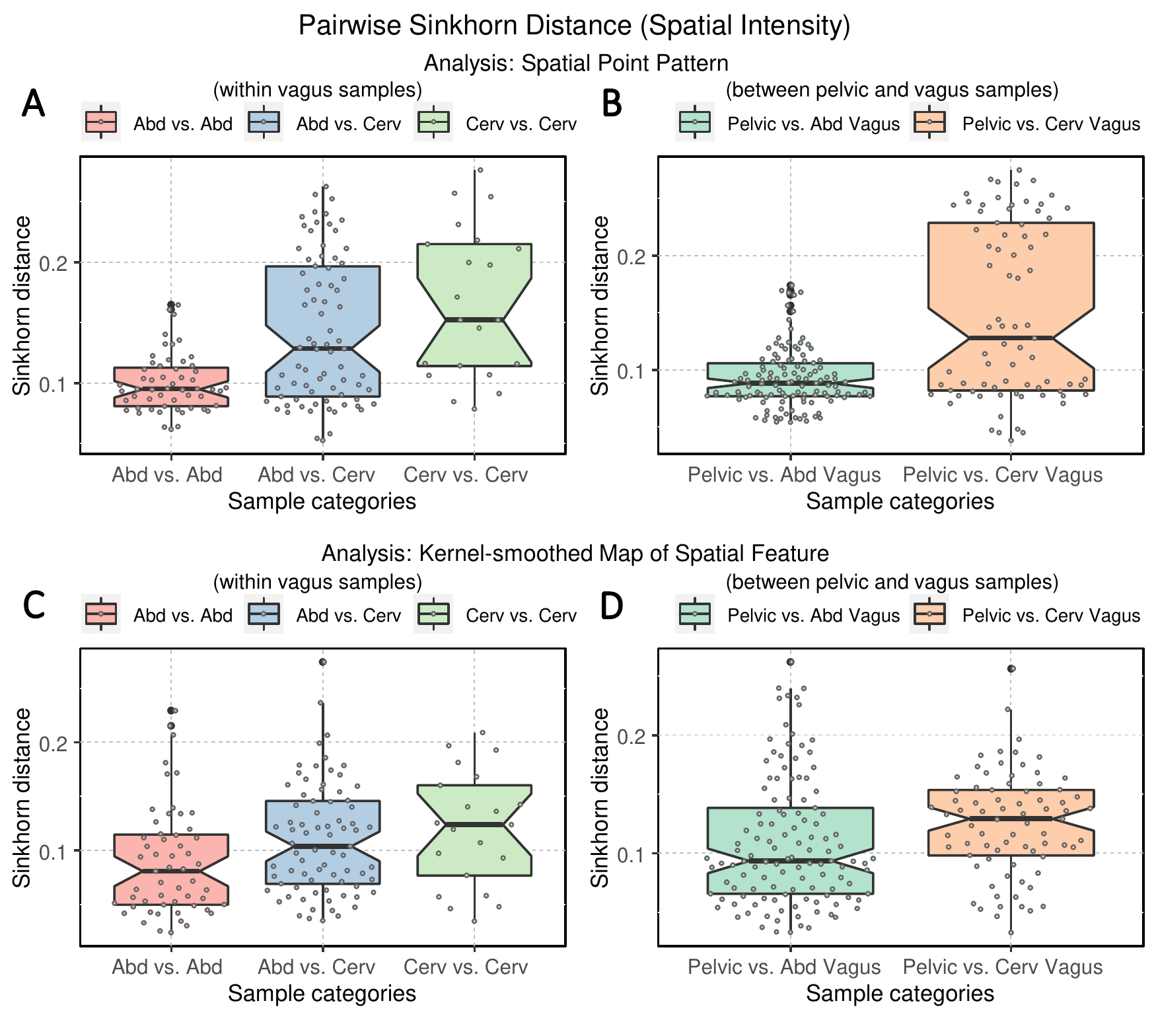}
    \caption{Boxplots displaying the ranges of Sinkhorn distance between the spatial intensity of the sub-categories of the nerve cross-sections - Abd: abdominal vagus, Cerv: cervical vagus, and pelvic. (A, B) Analysis of the spatial point patterns directly. (C, D) Analysis of the kernel-smoothed maps of the spatial features. The points show the individual Sinkhorn distances, revealing the hidden distribution.}
    \label{fig:t4_Bary_si_stat}
\end{figure}

We compute the Sinkhorn distances between every pair of nerve cross-sections for the four spatial features using spatial point patterns directly (in Section~\ref{subsec:exp1}) and kernel-smoothed bitmaps representation of the spatial features (in Section~\ref{subsec:exp2}). The resulting Sinkhorn embeddings are shown in Figure~\ref{fig:t4_si_1}, Figure~\ref{fig:t4_sf_1} and Figure~\ref{fig:bary_sf_1}. The created Sinkhorn space allows us to observe the similarities (or dissimilarities) of spatial intensities and second-order spatial properties.


The secondary statistical analysis performed on the embedded patterns generated by kernel-smoothing to mitigate the effects of the unequal number of axons revealed that the difference in spatial architecture between the vagus and pelvic nerves is relatively small (Mahalanobis distance $\Delta = 0.91$). However, the sample size is insufficient to determine whether this observed difference reflects biological reality or was the result of random chance. With $n_{\textrm{pelvic}} = 11$ and $n_{\textrm{vagus}} = 18$, the achieved power ($1-\beta$) is only $0.6$. To confirm the spatial architectural difference between vagus and pelvic nerve cross-sections, the required data set size should be at least $n = 26$ per class for $1-\beta=0.8$ and $\alpha=0.05$ in 2-D embedding, according to the collected preliminary results. In other words, any future research on the potential architectural difference between peripheral nerves' axonal organization (or modulation of such organization due to pathology or treatment) must use these preliminary effect sizes as a reasonable basis for necessary power analysis needed for experimental design. 

On the other hand, there is a substantial difference between the nerve cross-sections of males and females ($\Delta=1.246$, Hotelling T$^2$ test \textit{p}-value $=0.013$). However, this effect must be confirmed and replicated with an unconfounded sample set in which the correlation between sex and cross-section origin (pelvic vs. vagus) is absent. The result reported here is based on the assumption that there is indeed no statistically significant difference between pelvic and vagus. 

Regarding intraclass variability, vagus samples exhibit a significantly greater diversity of spatial architecture than pelvic samples when the raw spatial patterns are directly compared (Figure~\ref{fig:t4_si_1}B). However, this difference disappears when the kernel-smoothed spatial patterns are compared, indicating that it was driven mainly by the difference in the number of axons rather than the spatial architecture (Figure~\ref{fig:bary_sf_1} and \ref{fig:t4_Bary_si_stat_2}).


The vagus samples in our dataset are collected from the abdominal and cervical regions (see Table~\ref{tab:dataset}). Thus, looking into the degree of variability of the Sinkhorn distance within the sub-categories of the vagus samples and between the pelvic samples is helpful. Figure~\ref{fig:t4_Bary_si_stat} illustrates the range of Sinkhorn distance between the spatial intensity of every pair of nerve cross-sections (for both types of analysis), categorized as the following:
\begin{enumerate}
	\item within vagus samples
	\begin{enumerate}
		\item intra-class measurements (i) (abdominal vagus vs. abdominal vagus)
		\item intra-class measurements (ii) (cervical vagus vs. cervical vagus)
		\item inter-class measurements (abdominal vagus vs. cervical vagus)
	\end{enumerate}
	\item between pelvic and vagus samples
	\begin{enumerate}
		\item inter-class measurements (i) (pelvic vs. abdominal vagus)
		\item inter-class measurements (iii) (pelvic vs. cervical vagus)
	\end{enumerate}
\end{enumerate}

In Figure~\ref{fig:t4_Bary_si_stat}, we see the ranges of the Sinkhorn distances for the abovementioned sub-categories. The degree of variability is more prominent in the analysis of the raw spatial point patterns (Figure~\ref{fig:t4_Bary_si_stat}A and B) compared to the analysis of the kernel-smoothed bitmaps representing spatial features (Figure~\ref{fig:t4_Bary_si_stat}C and D). This observation applies not just to spatial intensity but also statistical second-order spatial statistics. Figure~\ref{fig:t4_Bary_si_stat} shows that variability within the abdominal vagus samples is substantially lower than the spatial variability within the cervical vagus cross-sections (\textit{p} $<0.001$). However, this notion has not been confirmed when looking at Figure~\ref{fig:t4_Bary_si_stat}C, which is based on pre-processing that eliminates the effect of axons' number. As before, this is most likely due to the low statistical power of the available sample size. The number of abdominal ($55$) and cervical ($21$) pairwise measurements is too small to confidently demonstrate the observed standardized effect size of $\Delta=0.59$. The required number of measurements for such effect size should be at least $45$ per class to achieve $1-\beta=0.8$ with $\alpha=0.05$. 

The much smaller standardized difference ($\Delta=0.3$) between two sites of vagus nerves sampling and pelvic nerves shown in Figure~\ref{fig:t4_Bary_si_stat}D can be confidently demonstrated due to the considerably larger number of available data points ($121$ pelvic vs. abd. vagus and $77$ pelvic vs. cervical vagus pairwise measurements). Therefore, we can state that the dissimilarity between pelvic and abdominal vagus spatial architectures is much smaller than the dissimilarity between pelvic and cervical vagus nerves (\textit{p} $= 0.0352$). In other words, abdominal vagus samples resemble pelvic cross-sections to a higher degree than cervical vagus cross-sections.
\section{Discussion}\label{sec:discussion}
While many modern feature learning methods can directly classify biological images based on structural differences, the critical issue is frequently not the ability to recognize differences but rather the capability to quantify the specific architectural aspects of biological structures in order to relate them to a function or the presence of some pathology. Neuroanatomy is one of the disciplines in which black-box image classifiers are undesirable because the goal of the research is to tie the image attributes to genuine anatomical and physiological knowledge regarding cell and tissue organization rather than merely sorting the images into pre-conceived classes. The research presented here provides an additional block in our multi-step sample and data processing pipeline, which also includes previously described data acquisition and image segmentation modules \citep{havton_human_2021, plebani_high-throughput_2022}.

We pursued the representation of the segmented unmyelinated axons in the TEM images of the peripheral nerve cross-sections as spatial point patterns not only to gain a better understanding of their neuroanatomy, but also to express the observed differences in a quantitative manner, which would enable a variety of automated analysis tasks in the future, such as automated image queries, image database retrieval, and biological image comparisons. While visual inspection of segmented images and their associated point patterns might provide some basic intuition regarding the spatial intensity, it is impossible to rely on the investigator's perceptual conceptions and judgments when examining more complex aspects such as local heterogeneous and anisotropic spatial features. Although global bulk measures of second-order spatial statistics, such as the $K$- and $L$-functions, help to represent and explore spatial interactions (randomness, inhibition, or clustering), they fail to capture local variations within biological structures. On the other hand, the local form of these spatial statistics functions generates yet another complex spatial pattern, leaving scientists with an equally tricky quantification problem. In this context, our analytical approach that captures differences between arrangements of any spatial distributions to form a visualizable embedding that enables straightforward comparison between complex structures provides a simple-to-use tool for neuroanatomists and computational neuroscientists.

There are two notable limitations associated with the demonstrated methods and their particular implementation. As previously indicated, the purported differences are relatively small, and even though they may be statistically significant, they could be biologically insubstantial. There is no reason to expect that the spatial arrangement would be drastically altered in samples that did not represent any recognized pathology. We realize that the value of the method would be illustrated in a more dramatic manner if the detected differences were related to a specific biological mechanistic model, particularly associated with a disease or an abnormality. On the other hand, it is essential to note that the ability to gather and evaluate a large comprehensive set of neuropathology data is contingent upon the availability of analytical tools. Therefore, the scarcity of labeled and segmented images is partially attributable to the absence of an analytic framework, casting doubt on the systematic benefit of acquiring comprehensive peripheral nerve data. We certainly hope that the conceptualization and presentation of our approach will encourage neuroanatomists to collect more data regarding peripheral nerves resulting in larger-scale quantitative anatomical studies. 

\begin{figure}
    \centering
    \includegraphics[width=\textwidth]{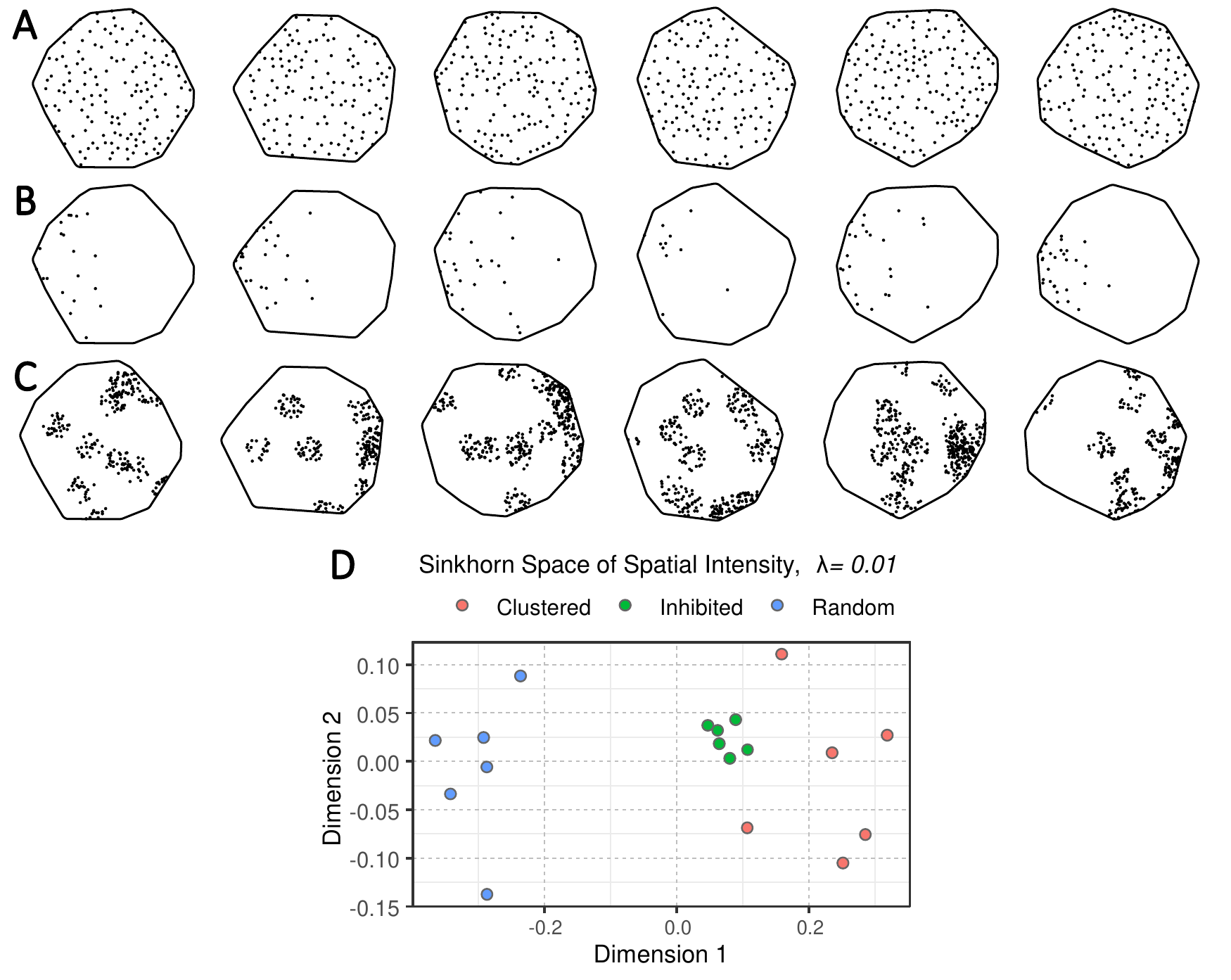}
    \caption{An illustration of spatial point patterns with different inhomogeneous spatial organizations. (A) Spatial inhibition. (B) Spatial randomness. (C) Spatial clustering. (D) An embedding of the spatial intensity of the point patterns shown in (A-C) in the Sinkhorn space. The inhibited, random, and clustered point patterns are shown in green, blue, and orange, respectively.}
    \label{fig:intro_ex}
\end{figure}

The second concern relates to the interpretability of the absolute position of patterns in the computed embeddings. Although Sinkhorn embedding produces 2-D maps of similarity, it may not be easy to interpret the resulting position of points in the Sinkhorn space. Fortunately, the proposed embedding can be coupled with data simulations, allowing for the production of any arrangement of biological properties within any shape of interest. Figure~\ref{fig:intro_ex}(A-C) depicts a toy example of 18 simulated spatial point patterns with spatial inhibition, randomness, and clustering, respectively. These point patterns are formed by inhomogeneous point processes, namely the hardcore process, the Poisson process, and the Matern cluster process \citep{baddeley2015spatial}. For the simulation of the toy example, we used a hardcore process with $400$ points per unit area and a hardcore distance of $0.04$ (the points are not allowed to be within $0.04$ unit of each other, ensuring inhibition). We used a Poisson process with inhomogeneous intensity function $\rho(x,y)$, shown in Equation~\ref{eq:poisson}, and a Matern cluster process with inhomogeneous intensity function $\rho(x,y)$ for the cluster centers, shown in Equation~\ref{eq:matclust}, cluster radius $0.10$ and $25$ points per cluster.
\begin{equation}\label{eq:poisson}
    \rho(x,y) = 100\times \exp(-5x)
\end{equation}
\begin{equation}\label{eq:matclust}
    \rho(x,y) = 10\times \exp(2\lvert x\rvert - 1)
\end{equation}
While the spatial structure of the point patterns from the nerve cross-sections is complex, consisting of inhomogeneous, anisotropic, and numerous spatial interactions, the simulated realizations are formed from a single point process and are relatively simple to explain. The embedding of the spatial intensity of the simulated point patterns in the Sinkhorn space is depicted in Figure~\ref{fig:intro_ex}D, where a clear distinction can be seen between samples with different spatial organizations. Similar simulations can be used for any type of spatial point pattern, providing a readily explainable, semi-mechanistic rationale for the emergence of the patterns and an interpretable representation of their properties and reasons for separation.

Although we focused here on unmyelinated axons, the computational pipeline is applicable to multi-type point patterns and spatial research outside of neuroscience. This work demonstrates that the spatial architecture of unmyelinated axons in peripheral nerve cross-sections is neither uniform nor random but forms complex and rich arrangements. In order to simulate such a complicated spatial form, hybrid point processes are required. In the future, we plan to focus on spatial modeling and further classification of peripheral nerve cross-sections. The similarity (or dissimilarity) measure we established in this study will be a foundation for these modeling tasks.
\section{Conclusions}
In this report, we examine one of the key research problems in neuroanatomy, the fundamental description, measurement, and quantification of the spatial arrangement of axons in peripheral nerves such as the vagus and pelvic nerves. This topic is significant not only from the basic neuroanatomical standpoint, but also due to the growing importance of peripheral nerve electrostimulation approaches, which rely on a precise understanding of the peripheral nerve architecture during the modeling and development phases \citep{Eiber2021}. We believe that quantitative analysis, comparisons, and visualization of spatial arrangement can provide valuable insight to neuroanatomists, computational neuroscientists, and engineers working in the field of electrostimulation. We also believe that the presented method can be easily adapted to other biological fields, including spatial proteomics and genomics \citep{Ji2020,Hickey2022}.

\section*{Conflict of Interest Statement}
The authors declare no competing interests.

\section*{Author Contributions}
BR conceived and planned the spatial analysis study; AP contributed to the mathematical models; EP and MD preprocessed the microscopy data and designed the segmentation pipeline; TP and JRK collected and processed the biological samples. TP provided neuroscience expertise and envisioned the quantitative peripheral nerve research; LAH and NB collected and annotated the microscopy images; DJ curated the data and performed image preprocessing and mosaicing; ASS and BR executed the study and co-wrote the manuscript with input from all the researchers.

\section*{Funding}

This work was supported by the National Institutes of Health, Office of the Director, Stimulating Peripheral Activity to Relieve Conditions (SPARC) Program under Award Number OT2OD023847 (TP, BR, DJ, MMD, EP, ASS), OT2OD023872 (JRK), and OT2OD026585 (LH, NB); and by the U.S. Department of Energy's Advanced Scientific Computing Research program through grant DE-SC-0022260 (AP, ASS). The content of this work is solely the responsibility of the authors and does not necessarily represent the official views of the National Institutes of Health or the Department of Energy.



\section*{Data Availability Statement}

Microscopy data associated with this study, were collected as part of the Stimulating Peripheral Activity to Relieve Conditions (SPARC) program and are available through the SPARC Portal (RRID: SCR\_017041) under a CC-BY 4.0 license.


\bibliographystyle{Frontiers-Harvard} 
\bibliography{paper-list}



\end{document}